\documentclass[11pt,a4paper]{article}

\usepackage{amsmath}
\usepackage{amssymb}
\usepackage{amsfonts}
\usepackage{amsthm}
\usepackage{graphicx}
\usepackage{float}
\usepackage{slashed}
\usepackage{multirow}
\usepackage{booktabs}
\usepackage{bm}
\usepackage{geometry}
\usepackage{cite}
\usepackage{xcolor}
\usepackage{hyperref}
\usepackage{subcaption}

\hypersetup{
colorlinks=true,
linkcolor=blue,
citecolor=blue,
urlcolor=blue
}

\title{
\bf
One-Zero Neutrino Textures and Resonant Type-II Leptogenesis:
Flavor-Resolved Thermal Evolution and Baryon Asymmetry
}

\author{
Avinanda Chaudhuri\\
\small Department of Physics,\\ Brahmananda Keshab Chandra College,\\ 111/2, B. T. Road, India - 700108\\
\small \thanks{\texttt{aviphys@gmail.com}}
}

\date{}

\begin{document}

\maketitle

\begin{abstract}

We investigate the viability of one-zero neutrino mass textures within the framework of resonant Type-II leptogenesis. Considering a two-triplet scalar realization of the Type-II seesaw mechanism, we analyze the compatibility between neutrino texture structures, CP asymmetry generation, and the observed baryon asymmetry of the Universe. We perform extensive numerical scans over the neutrino parameter space and classify the phenomenologically viable one-zero textures under both hierarchical and resonant leptogenesis scenarios.

We show that several one-zero textures remain compatible with neutrino oscillation data while simultaneously generating sizable CP asymmetries through resonant enhancement. We further investigate the thermal evolution of the generated asymmetry using Boltzmann equations and demonstrate the freeze-out behavior of the baryon asymmetry. Extending the analysis to a flavor-resolved framework, we study the separate evolution of electron, muon, and tau asymmetries and show that flavor-dependent washout effects play a crucial role in determining the final baryon asymmetry.

Our analysis establishes a direct connection between neutrino flavor textures, resonant thermal leptogenesis, and flavor-dependent baryogenesis dynamics.

\end{abstract}

\tableofcontents

\newpage

%===========================================================
\section{Introduction}
%===========================================================

The origin of neutrino masses and the observed baryon asymmetry of the Universe remain two of the most important open problems in particle physics and cosmology. Neutrino oscillation experiments~\cite{SuperK, SNO, KamLAND, DayaBay, T2K} have firmly established that neutrinos are massive and mixed, requiring an extension of the Standard Model. Simultaneously, cosmological observations indicate the existence of a tiny but non-vanishing baryon asymmetry~\cite{Planck2018},
\begin{equation}
Y_B^{\rm obs}
\simeq
8.7\times10^{-11},
\end{equation}
which cannot be explained within the Standard Model framework alone.

Among the various mechanisms proposed for neutrino mass generation, the Type-II seesaw mechanism~\cite{Minkowski, Yanagida, GellMann1979, Mohapatra, Schechter, Magg, Lazarides, MohapatraTypeII, Chun} provides a particularly elegant framework in which neutrino masses arise from the vacuum expectation value of scalar triplets. In its minimal realization, the Type-II seesaw introduces an $SU(2)_L$ scalar triplet whose Yukawa interactions directly generate Majorana neutrino masses after electroweak symmetry breaking.

Texture structures in the neutrino mass matrix provide an important approach for understanding the flavor structure of neutrino masses and mixing. In particular, one-zero textures have attracted considerable attention because they reduce the number of free parameters and can lead to predictive relations among neutrino observables~\cite{Frampton2002, Xing2002, GuoXing2003, Dev2007, Lashin2008, Meloni2012, LudlGrimus2014, Liao2014, Singh2007, Zhou2015, FritzschXing, GrimusLavoura, Branco2007, AltarelliFeruglio, KingLuhn}. Such textures can also reveal underlying flavor symmetries and constrain the allowed regions of the neutrino parameter space.

On the other hand, leptogenesis provides a compelling mechanism for generating the observed baryon asymmetry through CP-violating lepton number violating interactions in the early Universe. In conventional Type-II leptogenesis scenarios involving a single scalar triplet, the generated CP asymmetry is typically insufficient to explain the observed baryon asymmetry. However, in the presence of two nearly degenerate scalar triplets, resonant enhancement can significantly amplify the CP asymmetry and lead to successful baryogenesis~\cite{Fukugita, Davidson, Buchmuller, Nardi, Abada, Pilaftsis, PilaftsisUnderwood, Dev, Hambye, MaSarkar, Rossi}.

Motivated by these considerations, in this work we investigate the compatibility between one-zero neutrino mass textures and resonant Type-II leptogenesis. Our primary goal is to determine whether the flavor structures imposed by one-zero textures can simultaneously satisfy neutrino oscillation constraints and produce a viable baryon asymmetry through resonantly enhanced thermal leptogenesis.

The paper is organized as follows. In Sec.~2 we briefly review the Type-II seesaw framework and neutrino mass generation. In Sec.~3 we discuss the one-zero texture structures and their phenomenological constraints. Sec.~4 is devoted to resonant Type-II leptogenesis and CP asymmetry generation. In Sec.~5 we present the flavored Boltzmann evolution. Our numerical analysis and results are contained in Sec.~6. Finally, we summarize our conclusions in Sec.~7.

%===========================================================
\section{Type-II Seesaw Framework}
%===========================================================

\subsection{Scalar Triplet Extension of the Standard Model}

In the Type-II seesaw mechanism, neutrino masses are generated through the introduction of an $SU(2)_L$ scalar triplet with hypercharge $Y=2$. The scalar triplet can be written as
\begin{equation}
\Delta
=
\begin{pmatrix}
\Delta^+/\sqrt{2} & \Delta^{++} \\
\Delta^0 & -\Delta^+/\sqrt{2}
\end{pmatrix}.
\end{equation}
The interaction of the scalar triplet with the lepton doublets is described by the Yukawa Lagrangian
\begin{equation}
\mathcal{L}_Y
=
Y_{\alpha\beta}
\,
L_\alpha^T
C
i\sigma_2
\Delta
L_\beta
+
{\rm h.c.},
\end{equation}

where
\begin{equation}
\alpha,\beta=e,\mu,\tau,
\end{equation}
$L_\alpha$ denotes the left-handed lepton doublets, $C$ is the charge conjugation matrix, and $Y_{\alpha\beta}$ is a complex symmetric Yukawa coupling matrix.

The scalar potential contains the trilinear interaction
\begin{equation}
V
\supset
\mu
\,
H^T
i\sigma_2
\Delta^\dagger
H
+
{\rm h.c.},
\end{equation}
where $H$ denotes the Standard Model Higgs doublet. After electroweak symmetry breaking, the neutral component of the scalar triplet acquires an induced vacuum expectation value,
\begin{equation}
v_\Delta
\simeq
\frac{\mu v^2}{M_\Delta^2},
\end{equation}
where
\begin{equation}
v\simeq246~{\rm GeV}
\end{equation}
is the Higgs vacuum expectation value and $M_\Delta$ denotes the triplet mass scale. In this work we have assumed $M_\Delta \sim 10^{11}\ {\rm GeV}$, $\mu \sim 10^{7}\ {\rm GeV}$ and $v_\Delta \sim 10^{-1}\ {\rm eV}$.

The Majorana neutrino mass matrix is then generated as
\begin{equation}
M_\nu
=
\sqrt{2}
\,
Y
\,v_\Delta.
\label{eq:typeII_mass}
\end{equation}

Thus, within the Type-II seesaw framework, the neutrino mass matrix directly reflects the flavor structure of the triplet Yukawa couplings.

%===========================================================
\subsection{Two-Triplet Realization}
%===========================================================

In the minimal Type-II seesaw scenario involving a single scalar triplet, the generated CP asymmetry is generally insufficient for successful leptogenesis. To overcome this limitation, we consider an extension involving two scalar triplets~\cite{Chaudhuri2015, Chaudhuri2016, Chaudhuri2026},
\begin{equation}
\Delta_1,
\qquad
\Delta_2,
\end{equation}
with nearly degenerate masses,
\begin{equation}
M_1
\simeq
M_2.
\end{equation}

The Yukawa interactions are generalized as
\begin{equation}
\mathcal{L}_Y
=
Y^{(1)}_{\alpha\beta}
L_\alpha^T
C
i\sigma_2
\Delta_1
L_\beta
+
Y^{(2)}_{\alpha\beta}
L_\alpha^T
C
i\sigma_2
\Delta_2
L_\beta
+
{\rm h.c.}
\end{equation}

Similarly, the scalar potential contains two trilinear couplings,
\begin{equation}
V
\supset
\mu_1
H^T
i\sigma_2
\Delta_1^\dagger
H
+
\mu_2
H^T
i\sigma_2
\Delta_2^\dagger
H
+
{\rm h.c.}
\end{equation}

The effective neutrino mass matrix becomes
\begin{equation}
M_\nu
=
M_\nu^{(1)}
+
M_\nu^{(2)},
\end{equation}
where
\begin{equation}
M_\nu^{(i)}
=
\sqrt{2}
\,
Y^{(i)}
v_{\Delta_i}.
\end{equation}

The presence of two nearly degenerate triplets allows resonant enhancement of the CP asymmetry generated in triplet decays.

%===========================================================
\subsection{Neutrino Mixing and PMNS Matrix}
%===========================================================

The neutrino mass matrix is diagonalized through
\begin{equation}
U_{\rm PMNS}^T
M_\nu
U_{\rm PMNS}
=
{\rm diag}(m_1,m_2,m_3),
\label{pmns}
\end{equation}
where $m_i$ denote the physical neutrino masses and $U_{\rm PMNS}$ is the Pontecorvo-Maki-Nakagawa-Sakata mixing matrix~\cite{PMNS, MNS}.

We adopt the standard parametrization
\begin{equation}
U_{\rm PMNS}
=
\begin{pmatrix}
1 & 0 & 0 \\
0 & e^{i\alpha_1/2} & 0 \\
0 & 0 & e^{i\alpha_2/2}
\end{pmatrix}
U
(\theta_{12},
\theta_{23},
\theta_{13},
\delta),
\end{equation}
where $\theta_{12}$, $\theta_{23}$, and $\theta_{13}$ are the neutrino mixing angles, $\delta$ is the Dirac CP phase, and $\alpha_1,\alpha_2$ denote the Majorana phases.

The neutrino mass matrix in the flavor basis can therefore be reconstructed as
\begin{equation}
M_\nu
=
U_{\rm PMNS}
\,
{\rm diag}(m_1,m_2,m_3)
\,
U_{\rm PMNS}^T.
\label{eq:reconstructed_mass}
\end{equation}

This reconstructed neutrino mass matrix forms the basis of the one-zero texture analysis discussed in the following section.

%===========================================================
\section{One-Zero Texture Structures}
%===========================================================

Explicitly, the neutrino mass matrix can be written as
\begin{equation}
M_\nu
=
\begin{pmatrix}
M_{ee} & M_{e\mu} & M_{e\tau} \\
M_{e\mu} & M_{\mu\mu} & M_{\mu\tau} \\
M_{e\tau} & M_{\mu\tau} & M_{\tau\tau}
\end{pmatrix},
\end{equation}
where the matrix is symmetric due to the Majorana nature of neutrinos.

Texture structures correspond to imposing specific conditions on the entries of the neutrino mass matrix. In the present work we focus on one-zero textures in which a single independent matrix element vanishes or becomes strongly suppressed.

%===========================================================
\subsection{Classification of One-Zero Textures}
%===========================================================

For a symmetric $3\times3$ Majorana neutrino mass matrix, there exist six independent one-zero textures:
\begin{align}
M_{ee} &=0,
&
M_{e\mu} &=0,
&
M_{e\tau} &=0,
\nonumber\\[2mm]
M_{\mu\mu} &=0,
&
M_{\mu\tau} &=0,
&
M_{\tau\tau} &=0.
\end{align}

Each texture imposes a non-trivial complex constraint on the neutrino parameter space. The texture condition can be expressed generally as
\begin{equation}
M_{\alpha\beta}
=
\sum_{i=1}^{3}
U_{\alpha i}
U_{\beta i}
m_i
=
0,
\label{eq:texture_condition}
\end{equation}
where $U_{\alpha i}$ are the elements of the PMNS matrix.

Equation~(\ref{eq:texture_condition}) leads to correlations among:neutrino masses,  mixing angles,  Dirac CP phase and Majorana phases. Consequently, one-zero textures can significantly restrict the allowed neutrino parameter space and generate predictive flavor structures.

%===========================================================
\subsection{Texture Constraints and Oscillation Parameters}
%===========================================================

The neutrino oscillation parameters used in our analysis are taken within their experimentally allowed ranges~\cite{PDG, NuFIT}. In particular, we vary:
\begin{align}
\theta_{12},
\qquad
\theta_{23},
\qquad
\theta_{13},
\qquad
\delta,
\end{align}
together with the Majorana phases
\begin{equation}
\alpha_1,
\qquad
\alpha_2,
\end{equation}
and the lightest neutrino mass.

The neutrino mass eigenvalues are determined through the measured mass-squared differences:
\begin{align}
\Delta m_{21}^2
&=
m_2^2-m_1^2,
\\[2mm]
\Delta m_{31}^2
&=
m_3^2-m_1^2
\qquad
{\rm (NH)},
\\[2mm]
\Delta m_{32}^2
&=
m_3^2-m_2^2
\qquad
{\rm (IH)}.
\end{align}

For each point in parameter space, the reconstructed neutrino mass matrix is tested against the one-zero texture conditions. Since exact texture zeros are difficult to realize numerically, we instead impose the condition that a given matrix element remains sufficiently suppressed:
\begin{equation}
|M_{\alpha\beta}|
<
\varepsilon_{\rm tex},
\end{equation}
where $\varepsilon_{\rm tex} \sim 10^{-3}\ {\rm eV}$ denotes the texture threshold.

In our numerical analysis we scan over the neutrino parameter space and identify the minimum achievable value of each texture element. This allows us to classify which one-zero textures are naturally compatible with neutrino oscillation data.

%===========================================================
\subsection{Connection to Flavor Geometry}
%===========================================================

An important aspect of one-zero textures is their direct connection to the flavor geometry of the neutrino sector~\cite{Blanchet, Beneke, Cirigliano2009, Garny2009, Kartavtsev2016, Blanchet2010, Deppisch2011}. Since the Type-II seesaw neutrino mass matrix is directly proportional to the triplet Yukawa matrix,
\begin{equation}
M_\nu
\propto
Y,
\end{equation}
the texture structures strongly constrain the flavor couplings responsible for leptogenesis.

As a result, the texture conditions influence:
\begin{itemize}
\item the magnitude of the CP asymmetry,
\item flavor-dependent washout effects,
\item and the thermal survival of the generated asymmetry.
\end{itemize}

This establishes a direct link between neutrino flavor textures and thermal baryogenesis, providing the central phenomenological framework explored in this work.

%===========================================================
\subsection{Texture Constraints and CP-Phase Correlations}
%===========================================================

An important consequence of one-zero texture conditions is the emergence of non-trivial correlations among the CP phases of the PMNS matrix.

As an illustrative example, consider the texture condition
\begin{equation}
M_{e\mu}=0.
\end{equation}

Using Eq.(\ref{pmns}) the texture condition becomes
\begin{equation}
m_1U_{e1}U_{\mu1}
+
m_2U_{e2}U_{\mu2}
+
m_3U_{e3}U_{\mu3}
=
0.
\label{eq:memu_exact}
\end{equation}

Substituting the PMNS matrix elements and expanding to leading order in \(s_{13}\), we obtain
\begin{equation}
m_1c_{12}s_{12}c_{23}
+
m_2s_{12}c_{12}c_{23}e^{i\alpha_1}
+
m_3s_{13}s_{23}e^{i(\alpha_2-\delta)}
\simeq
0.
\label{eq:memu_phase}
\end{equation}

Equation~(\ref{eq:memu_phase}) explicitly demonstrates that the texture condition requires a non-trivial interference among the neutrino mass eigenstates and CP phases.

Rearranging Eq.~(\ref{eq:memu_phase}) gives an approximate relation for the Dirac CP phase:
\begin{equation}
e^{-i\delta}
\sim
\frac{
c_{12}s_{12}c_{23}
}{
m_3s_{13}s_{23}
}
\left(
m_1+m_2e^{i\alpha_1}
\right)
e^{-i\alpha_2}.
\end{equation}

Taking the imaginary part yields the approximate phase correlation
\begin{equation}
\tan\delta
\sim
\frac{
m_2\sin\alpha_1
}{
m_1+m_2\cos\alpha_1
}.
\label{eq:tan_delta}
\end{equation}

Equation~(\ref{eq:tan_delta}) shows that the one-zero texture condition naturally leads to preferred regions in the Dirac CP phase space.

\begin{figure}[t!]
\centering
\includegraphics[width=0.82\textwidth]
{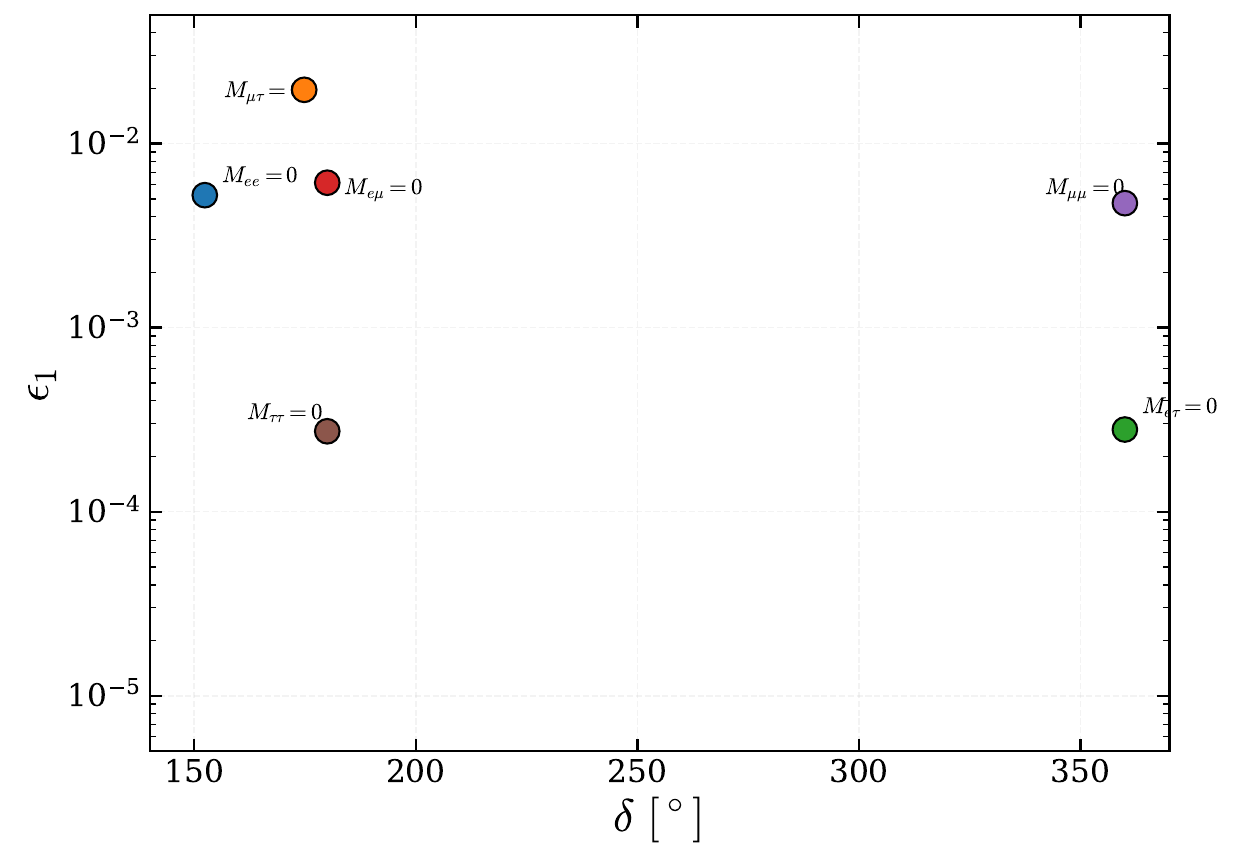}
\caption{
Distribution of the CP asymmetry parameter $\epsilon_1$ as a function of the Dirac CP phase $\delta$ for the different one-zero neutrino texture structures.
Each point corresponds to the minimum achievable texture configuration obtained from the numerical scan.
Distinct clustering patterns emerge for different textures, reflecting the flavor geometry imposed by the one-zero texture conditions.
The figure demonstrates that even in the resonantly enhanced regime, the generated CP asymmetry retains non-trivial sensitivity to the underlying neutrino flavor structure and CP phase correlations.
}
\label{fig:phase_distribution}
\end{figure}

In particular, destructive interference among the neutrino mass eigenstates favors phase alignments near
\begin{equation}
\delta
\simeq
\pi,
\end{equation}
consistent with the phase clustering patterns observed in Fig.~\ref{fig:phase_distribution}.

%===========================================================
% FIGURE 8
%===========================================================

This relation provides an analytic explanation for the emergence of texture-dependent phase structures in the resonant leptogenesis framework.

%===========================================================
\section{Resonant Type-II Leptogenesis}
%===========================================================

\subsection{Triplet Decays and Lepton Asymmetry}

In the Type-II seesaw framework, the scalar triplets can decay through the channels
\begin{align}
\Delta_i
&\rightarrow
L_\alpha
L_\beta,
\\[2mm]
\Delta_i
&\rightarrow
HH,
\end{align}
where $L_\alpha$ denotes the lepton doublets and $H$ represents the Standard Model Higgs doublet.

These decays violate lepton number and, in the presence of CP-violating phases, generate a net lepton asymmetry in the early Universe. The generated asymmetry is subsequently converted into baryon asymmetry through electroweak sphaleron processes.

The decay width of the scalar triplets thus can be written as:
\begin{equation}
\Gamma_i
=
\Gamma(\Delta_i\rightarrow LL)
+
\Gamma(\Delta_i\rightarrow HH).
\end{equation}

The leptonic decay width is given approximately by
\begin{equation}
\Gamma(\Delta_i\rightarrow LL)
=
\frac{
M_i
}{
8\pi
}
{\rm Tr}
\left(
Y^{(i)}
Y^{(i)\dagger}
\right),
\end{equation}
while the Higgs decay width is
\begin{equation}
\Gamma(\Delta_i\rightarrow HH)
=
\frac{
|\mu_i|^2
}{
8\pi M_i
}.
\end{equation}

The total decay width therefore depends directly on the triplet Yukawa structure and the trilinear scalar couplings.

%===========================================================
\subsection{CP Asymmetry Generation}
%===========================================================

The CP asymmetry arises through the interference between the tree-level decay amplitudes and the one-loop self-energy contributions involving the second scalar triplet. The flavored CP asymmetry associated with the decay of $\Delta_1$ can be expressed schematically as
\begin{equation}
\epsilon_1
=
\frac{
\Gamma(\Delta_1\rightarrow LL)
-
\Gamma(\Delta_1^\dagger\rightarrow \bar L\bar L)
}{
\Gamma_1
}.
\end{equation}

%===========================================================

%===========================================================
\begin{figure}[t!]
\centering
\includegraphics[width=0.74\textwidth]
{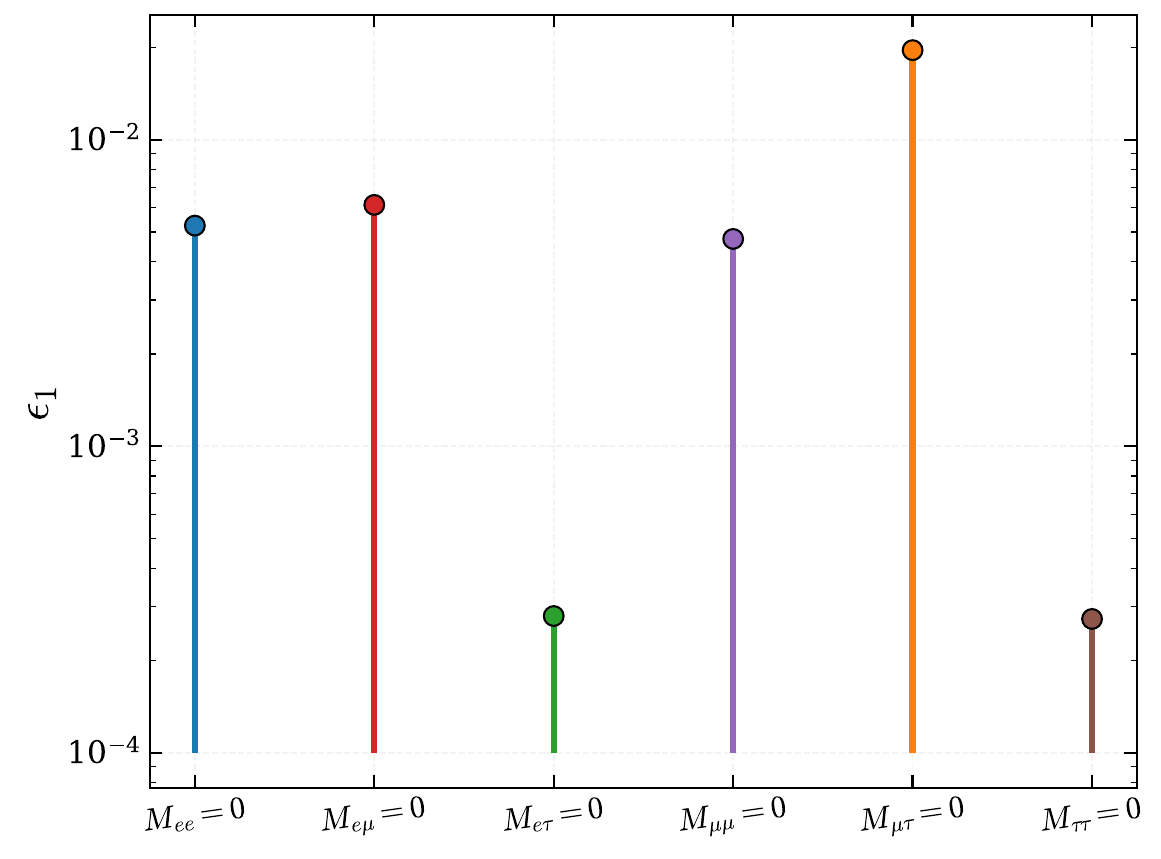}
\caption{
Maximum CP asymmetry generated for the different one-zero neutrino texture structures within the resonant Type-II seesaw framework.
The off-diagonal textures generally allow comparatively larger CP asymmetries due to the stronger interference among flavor contributions in the triplet Yukawa sector.
The figure illustrates the direct sensitivity of leptogenesis to the underlying neutrino flavor geometry.
}
\label{fig:cp_asymmetry}
\end{figure}
%===========================================================

In the two-triplet scenario, the dominant contribution originates from the self-energy diagram, leading to
\begin{equation}
\epsilon_1
\simeq
\frac{
1
}{
8\pi
}
\frac{
{\rm Im}
\left[
\mu_1
\mu_2^\ast
{\rm Tr}
\left(
Y^{(1)}
Y^{(2)\dagger}
\right)
\right]
}{
{\rm Tr}
\left(
Y^{(1)}
Y^{(1)\dagger}
\right)
M_1^2
+
|\mu_1|^2
}
R,
\label{eq:cp_asymmetry}
\end{equation}
where $R$ denotes the resonant enhancement factor.

The structure of Eq.~(\ref{eq:cp_asymmetry}) demonstrates that the CP asymmetry is highly sensitive to:
\begin{itemize}
\item the flavor structure of the Yukawa matrices,
\item the complex CP phases,
\item and the mass splitting between the triplets.
\end{itemize}

Consequently, the one-zero texture conditions can strongly affect the efficiency of leptogenesis.

\subsection{Texture-Dependent Structure of the CP Asymmetry}

An important feature of the Type-II seesaw framework is that the neutrino mass matrix is directly proportional to the triplet Yukawa couplings:
\begin{equation}
M_\nu^{(i)}
=
\sqrt{2}
Y^{(i)}
v_{\Delta_i}.
\end{equation}

Using
\begin{equation}
Y^{(i)}
=
\frac{
M_\nu^{(i)}
}{
\sqrt{2}v_{\Delta_i}
},
\end{equation}
the resonant CP asymmetry can be rewritten directly in terms of the neutrino mass matrices:
\begin{equation}
\epsilon_1
=
\frac{1}{16\pi}
\frac{
\mathrm{Im}
\left[
\mu_1\mu_2^*
\mathrm{Tr}
\left(
M_\nu^{(1)}
M_\nu^{(2)\dagger}
\right)
\right]
}{
v_{\Delta_1}v_{\Delta_2}
\left[
\frac{M_1^2}{2v_{\Delta_1}^2}
\mathrm{Tr}
\left(
M_\nu^{(1)}
M_\nu^{(1)\dagger}
\right)
+
|\mu_1|^2
\right]
}
R,
\label{eq:texture_exact}
\end{equation}
where \(R\) denotes the resonant enhancement factor.

The one-zero texture condition
\begin{equation}
(M_\nu)_{\alpha\beta}=0
\end{equation}
imposes the constraint
\begin{equation}
\sum_i
U_{\alpha i}
U_{\beta i}
m_i
=
0.
\end{equation}

Expanding the trace structure in Eq.~(\ref{eq:texture_exact}) gives
\begin{equation}
\mathrm{Tr}
\left(
M_\nu^{(1)}
M_\nu^{(2)\dagger}
\right)
=
\sum_{\rho,\sigma}
(M_\nu^{(1)})_{\rho\sigma}
(M_\nu^{(2)})_{\rho\sigma}^*.
\end{equation}

The texture condition therefore removes one flavor component from the interference structure entering the CP asymmetry. As a result, the asymmetry acquires a direct texture dependence:
\begin{equation}
\epsilon_1^{(\alpha\beta)}
\propto
\frac{1}{\Delta M}
\,
\mathrm{Im}
\left[
\sum_{(\rho\sigma)\neq(\alpha\beta)}
(M_\nu^{(1)})_{\rho\sigma}
(M_\nu^{(2)})_{\rho\sigma}^*
\right].
\label{eq:texture_epsilon}
\end{equation}

Equation~(\ref{eq:texture_epsilon}) demonstrates explicitly that the resonantly enhanced CP asymmetry retains direct sensitivity to the underlying neutrino flavor geometry even in the quasi-degenerate regime.

This analytic relation explains the texture-dependent clustering patterns observed in the numerical phase-space analysis and provides a direct connection between one-zero flavor structures and thermal leptogenesis.

%===========================================================
\subsection{Resonant Enhancement Mechanism}
%===========================================================

The resonant enhancement factor is given approximately by
\begin{equation}
R
=
\frac{
(M_2^2-M_1^2)
M_1\Gamma_2
}{
(M_2^2-M_1^2)^2
+
M_1^2\Gamma_2^2
}.
\end{equation}

Introducing the dimensionless mass splitting parameter
\begin{equation}
\Delta M
=
\frac{
M_2-M_1
}{
M_1
},
\end{equation}
the resonant condition corresponds to

\begin{equation}
\Delta M
\ll
1.
\end{equation}

Near the resonant regime,
\begin{equation}
M_2-M_1
\sim
\Gamma_2,
\end{equation}
the denominator of the resonant factor becomes strongly suppressed, leading to a substantial enhancement of the CP asymmetry.

To leading order, the asymmetry scales approximately as
\begin{equation}
\epsilon_1
\propto
\frac{1}{\Delta M}.
\label{eq:resonant_scaling}
\end{equation}

This inverse scaling relation forms the basis of the resonant enhancement mechanism explored numerically in this work.

%===========================================================
\section{Boltzmann Evolution and Flavor Dynamics}
%===========================================================
The thermal history is governed by a coupled system of Boltzmann equations~\cite{KolbTurner, Giudice} describing: the production and decay of the scalar triplets, inverse decay processes and flavor-dependent washout effects.   The Boltzmann analysis therefore provides the dynamical connection between the resonantly enhanced CP asymmetry and the final surviving baryon asymmetry.

%===========================================================
\subsection{Thermal Evolution Equations}
%===========================================================

We define the dimensionless inverse temperature parameter
\begin{equation}
z
=
\frac{M_1}{T},
\end{equation}
where \(M_1\) denotes the mass of the lighter scalar triplet and \(T\) is the temperature of the thermal plasma.

The Boltzmann equation governing the triplet abundance is given by
\begin{equation}
\frac{dY_\Delta}{dz}
=
-
D(z)
\left(
Y_\Delta
-
Y_\Delta^{\rm eq}
\right),
\label{eq:boltzmann_triplet}
\end{equation}
where:  \(Y_\Delta\) is the comoving triplet abundance, \(Y_\Delta^{\rm eq}\) denotes the equilibrium abundance and \(D(z)\) represents the thermally averaged decay function.

The evolution of the total \(B-L\) asymmetry is governed by
\begin{equation}
\frac{dY_{B-L}}{dz}
=
-
\epsilon_1
D(z)
\left(
Y_\Delta
-
Y_\Delta^{\rm eq}
\right)
-
W(z)
Y_{B-L},
\label{eq:boltzmann_bl}
\end{equation}
where:  \(\epsilon_1\) is the resonantly enhanced CP asymmetry  and \(W(z)\) denotes the washout kernel arising from inverse decay and lepton-number-violating scattering processes.

These coupled equations determine the complete thermal evolution of the generated asymmetry in the early Universe.

%===========================================================
\subsection{Decay and Washout Functions}
%===========================================================

The decay function is defined as
\begin{equation}
D(z)
=
K z
\frac{K_1(z)}{K_2(z)},
\end{equation}
where:
\begin{itemize}
\item \(K_1\) and \(K_2\) are modified Bessel functions,
\item and
\begin{equation}
K
=
\frac{\Gamma_\Delta}{H(T=M_1)}
\end{equation}
is the decay parameter.
\end{itemize}

The Hubble expansion rate in the radiation-dominated epoch is given by
\begin{equation}
H(T)
=
1.66
\sqrt{g_*}
\frac{T^2}{M_{\rm Pl}},
\end{equation}
where: \(g_*\) denotes the effective number of relativistic degrees of freedom
 and \(M_{\rm Pl}\) is the Planck mass.

The equilibrium triplet abundance is approximated by
\begin{equation}
Y_\Delta^{\rm eq}(z)
=
\frac{45}{4\pi^4 g_*}
z^2
K_2(z).
\end{equation}

Similarly, the washout kernel is parametrized as
\begin{equation}
W(z)
=
\frac{1}{2}
K z^3 K_1(z),
\end{equation}
which efficiently suppresses the generated asymmetry at high temperatures.

The competition between the decay and washout terms determines whether the generated asymmetry survives thermal equilibration.

%===========================================================
\subsection{Flavor-Resolved Thermal Evolution}
%===========================================================

Since the one-zero textures intrinsically constrain the flavor structure of the neutrino sector, it is important to investigate the flavored evolution of the generated asymmetry.

We therefore extend the analysis by separately evolving the flavor asymmetries:
\begin{equation}
Y_{\Delta_e},
\qquad
Y_{\Delta_\mu},
\qquad
Y_{\Delta_\tau}.
\end{equation}

The flavored Boltzmann equations can be written schematically as
\begin{equation}
\frac{dY_{\Delta_\alpha}}{dz}
=
-
\epsilon_\alpha
D(z)
\left(
Y_\Delta-Y_\Delta^{\rm eq}
\right)
-
W_\alpha(z)
Y_{\Delta_\alpha},
\end{equation}

where the washout term $W_\alpha(z)$ depends on the individual flavor couplings.

\begin{figure}[t]
\centering
\includegraphics[width=0.84\textwidth]
{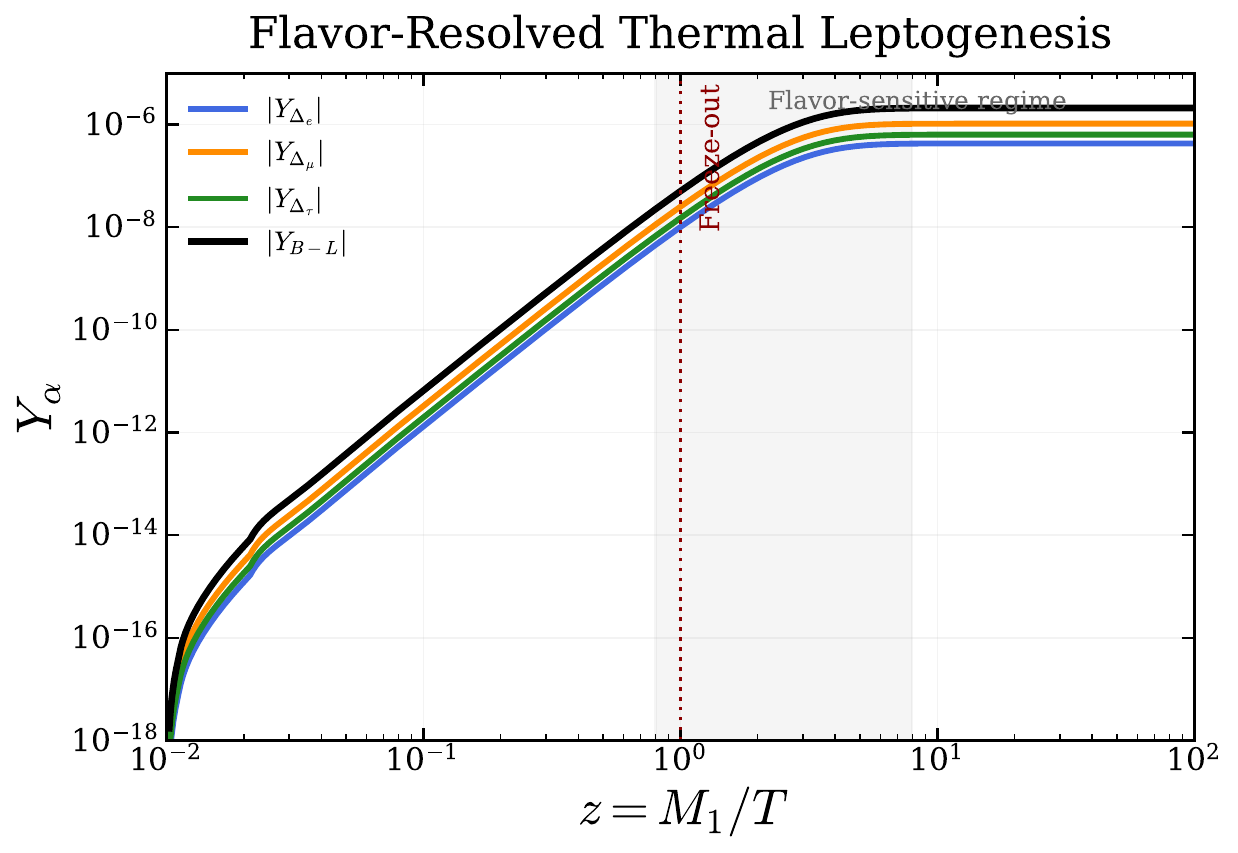}
\caption{
Flavor-resolved Boltzmann evolution of the generated lepton asymmetries in the resonant Type-II leptogenesis framework.
The individual flavor sectors undergo different thermal histories due to flavor-dependent CP asymmetries and washout effects.
A hierarchical structure emerges among the surviving asymmetries, demonstrating the direct impact of neutrino flavor geometry on thermal leptogenesis.
}
\label{fig:flavor_evolution}
\end{figure}

The resulting flavored evolution is shown in Fig.~\ref{fig:flavor_evolution}.

The numerical evolution reveals a hierarchical structure among the flavor asymmetries:
\begin{equation}
|Y_{\Delta_\mu}|
>
|Y_{\Delta_\tau}|
>
|Y_{\Delta_e}|.
\end{equation}

This hierarchy originates from the flavor-dependent washout efficiencies and the structure of the CP asymmetries.

The flavored analysis therefore establishes a direct connection between:
\begin{itemize}
\item neutrino texture geometry,
\item flavor-dependent washout,
\item and thermal baryogenesis.
\end{itemize}

This provides one of the central phenomenological results of the present work.

The numerical solution of the flavored Boltzmann equations is shown in Fig.~\ref{fig:flavor_evolution}.

%===========================================================
%===========================================================

Figure~\ref{fig:flavor_evolution} demonstrates that the flavored asymmetries do not evolve identically.
The muon asymmetry survives most efficiently against washout effects, while the electron asymmetry experiences comparatively stronger suppression.

This behavior is consistent with the texture-dependent flavored CP asymmetry structure derived analytically in the previous section and provides a direct dynamical realization of flavor-dependent baryogenesis.

%===========================================================
\subsection{Freeze-Out Dynamics}
%===========================================================

Initially, the scalar triplets remain in thermal equilibrium with the primordial plasma:
\begin{equation}
Y_\Delta
\simeq
Y_\Delta^{\rm eq}.
\end{equation}

As the temperature drops below the triplet mass scale,
\begin{equation}
T
<
M_1,
\end{equation}
the decay rate becomes insufficient to maintain equilibrium and the triplets gradually freeze out.

The generated lepton asymmetry simultaneously undergoes partial washout through inverse decay and scattering processes before approaching a constant asymptotic value.

\begin{figure}[t]
\centering
\includegraphics[width=0.84\textwidth]
{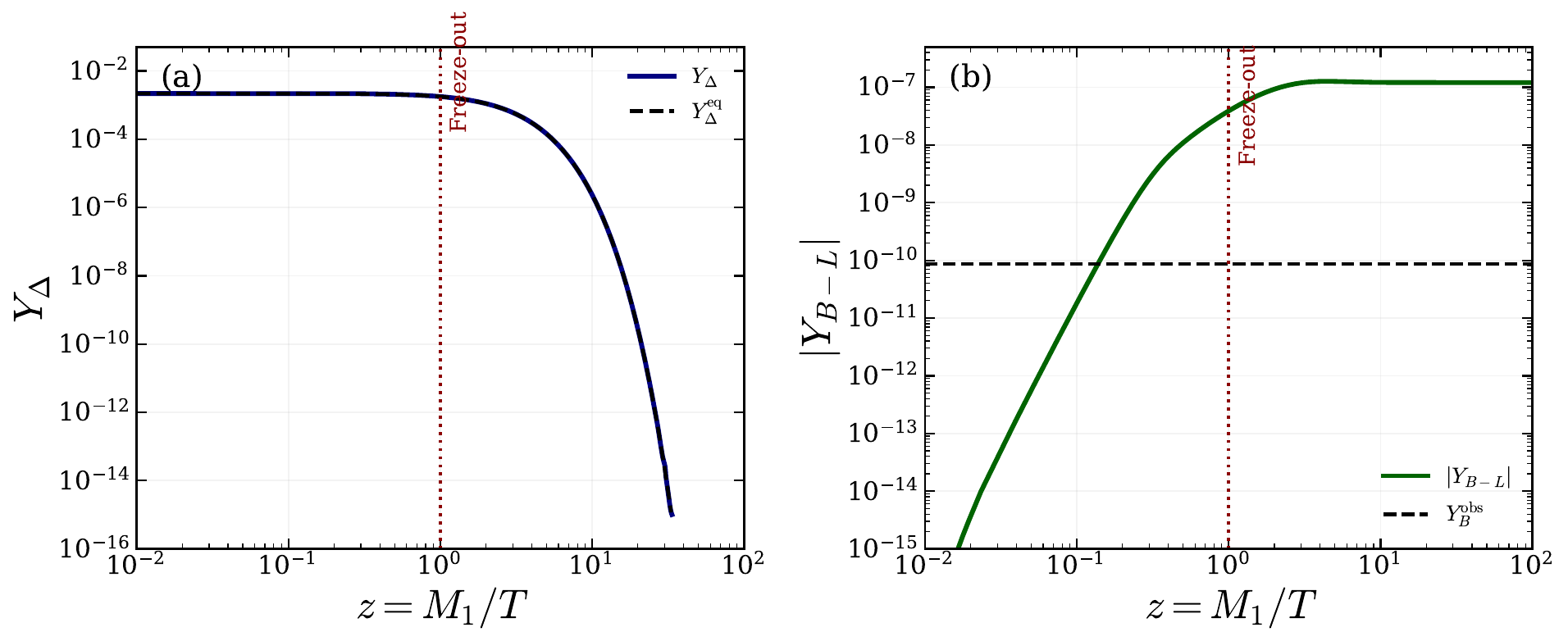}
\caption{
Thermal freeze-out evolution of the scalar triplet abundance and the generated baryon asymmetry.
Initially, the triplet abundance follows the equilibrium thermal distribution.
As the temperature drops below the triplet mass scale, the triplets decouple from the plasma and the baryon asymmetry approaches a constant late-time value after washout processes become inefficient.
}
\label{fig:freezeout}
\end{figure}

The numerical freeze-out evolution obtained from the Boltzmann equations is illustrated in Fig.~\ref{fig:freezeout}.

%===========================================================
%===========================================================

As shown in Fig.~\ref{fig:freezeout}, the scalar triplets closely track the equilibrium distribution at early times before progressively decoupling from the thermal plasma.

Simultaneously, the generated baryon asymmetry freezes into a stable asymptotic value once the washout interactions become sufficiently weak.

Successful baryogenesis therefore requires a delicate interplay among:
\begin{itemize}
\item CP asymmetry generation,
\item resonant enhancement,
\item flavor-dependent washout,
\item and thermal freeze-out dynamics.
\end{itemize}

%===========================================================
\subsection{Conversion to Baryon Asymmetry}
%===========================================================

The final baryon asymmetry is related to the surviving \(B-L\) asymmetry through sphaleron conversion~\cite{Kuzmin1985, Khlebnikov1988, HarveyTurner1990, Burnier2006, ArnoldMcLerran, Shaposhnikov1987}:
\begin{equation}
Y_B
=
\frac{28}{79}
Y_{B-L}.
\end{equation}

The observed baryon asymmetry measured by cosmological observations is
\begin{equation}
Y_B^{\rm obs}
\simeq
8.7\times10^{-11}.
\end{equation}

This observed value serves as the primary phenomenological constraint throughout the present analysis.

The Boltzmann evolution therefore provides the dynamical bridge connecting:
\begin{equation}
\text{Texture Geometry}
\longrightarrow
\text{Flavor-Dependent CP Asymmetry}
\longrightarrow
\text{Thermal Evolution}
\longrightarrow
Y_B.
\end{equation}

This establishes the central physical framework explored in the present work.

%===========================================================
\section{Numerical Analysis and Discussion}
%===========================================================

In order to investigate the compatibility between one-zero neutrino textures and resonant Type-II leptogenesis, we perform extensive numerical scans over the neutrino parameter space. The analysis includes both the neutrino oscillation parameters and the resonant mass splitting parameter controlling the enhancement of the CP asymmetry.

For each scan point we reconstruct the neutrino mass matrix using Eq.~(\ref{eq:reconstructed_mass}) and evaluate the corresponding one-zero texture conditions. Simultaneously, the CP asymmetry and baryon asymmetry are computed within the resonant Type-II leptogenesis framework.

%===========================================================
\begin{table}[t!]
\centering
\caption{
Representative benchmark parameters used in the numerical analysis of resonant Type-II leptogenesis with one-zero neutrino textures.
The chosen benchmark corresponds to the resonantly enhanced regime capable of reproducing the observed baryon asymmetry while maintaining strong texture suppression.
}
\label{tab:benchmark}
\renewcommand{\arraystretch}{1.2}
\begin{tabular}{cc}
\hline\hline
Parameter & Benchmark Value \\
\hline

Triplet mass scale $M_1$ 
& $10^{11}\ {\rm GeV}$ \\

Mass splitting parameter $\Delta M$
& $10^{-5}$ \\

Triplet vacuum expectation value $v_\Delta$
& $10^{-1}\ {\rm eV}$ \\

Trilinear coupling $\mu$
& $10^{7}\ {\rm GeV}$ \\

Texture threshold $\epsilon_{\rm tex}$
& $10^{-3}\ {\rm eV}$ \\

CP asymmetry $\epsilon_1$
& $\mathcal{O}(10^{-3})$ \\

Final baryon asymmetry $Y_B$
& $\sim 8.7\times10^{-11}$ \\

Light neutrino mass scale
& $\sim 0.05\ {\rm eV}$ \\

\hline\hline
\end{tabular}
\end{table}
%===========================================================

%===========================================================
\subsection{Texture Viability and Hierarchy}
%===========================================================

We first investigate the minimum achievable suppression for each one-zero texture. The numerical scans reveal a clear hierarchy among the six independent textures.

The off-diagonal textures
\begin{equation}
M_{e\mu}=0,
\qquad
M_{e\tau}=0,
\end{equation}
are found to be significantly easier to realize compared to the diagonal textures
\begin{equation}
M_{\mu\mu}=0,
\qquad
M_{\tau\tau}=0.
\end{equation}

%===========================================================
\begin{figure}[t!]
\centering
\includegraphics[width=0.72\textwidth]
{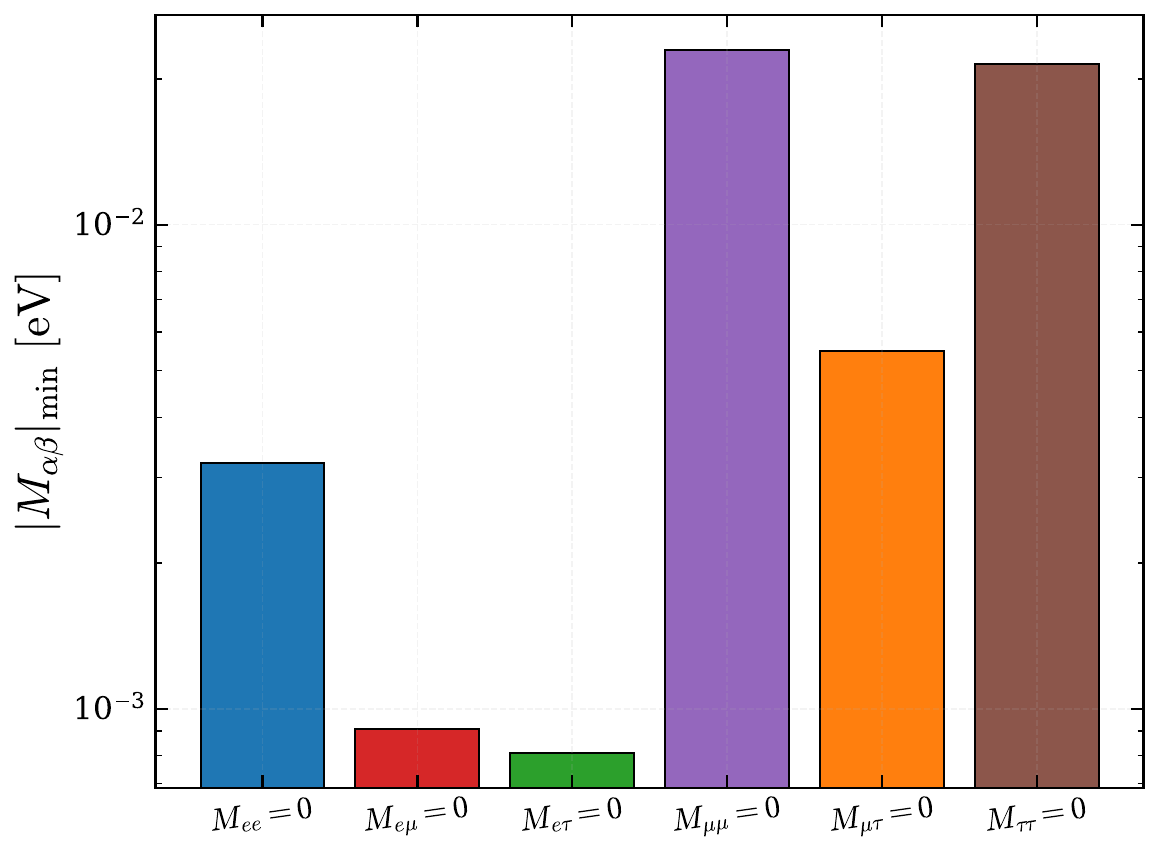}
\caption{
Hierarchy of the minimum achievable one-zero texture values obtained from the numerical parameter scan. The off-diagonal textures are generally more compatible with oscillation data than the diagonal texture structures.
}
\label{fig:texture_hierarchy}
\end{figure}
%===========================================================

This behavior as shown in Fig.(\ref{fig:texture_hierarchy}) originates from the interference structure of the PMNS matrix elements entering the texture conditions. The off-diagonal entries allow stronger cancellations among the neutrino mass eigenstate contributions, while the diagonal textures generally require more restrictive parameter tuning.

The texture hierarchy obtained numerically therefore reflects the underlying flavor geometry of the neutrino sector.

An important consequence of this result is that the texture structure directly affects the allowed Yukawa couplings entering the leptogenesis framework. Since the Type-II seesaw neutrino mass matrix is proportional to the triplet Yukawa matrix, the texture conditions strongly constrain the flavor structure responsible for CP asymmetry generation.

%===========================================================
\subsection{Failure of Hierarchical Type-II Leptogenesis}
%===========================================================

Our analysis demonstrates that hierarchical Type-II leptogenesis fails to generate the observed baryon asymmetry for the one-zero textures considered in this work.

In the hierarchical regime,
\begin{equation}
M_2
\gg
M_1,
\end{equation}
the self-energy contribution remains strongly suppressed, leading to small CP asymmetries:
\begin{equation}
\epsilon_1
\ll
1.
\end{equation}

Consequently, the resulting baryon asymmetry remains significantly below the observed cosmological value,
\begin{equation}
Y_B^{\rm obs}
\simeq
8.7\times10^{-11}.
\end{equation}
%
%===========================================================
\begin{figure*}[t!]
\centering

\begin{subfigure}[t]{0.42\textwidth}
\centering
\includegraphics[width=\textwidth]
{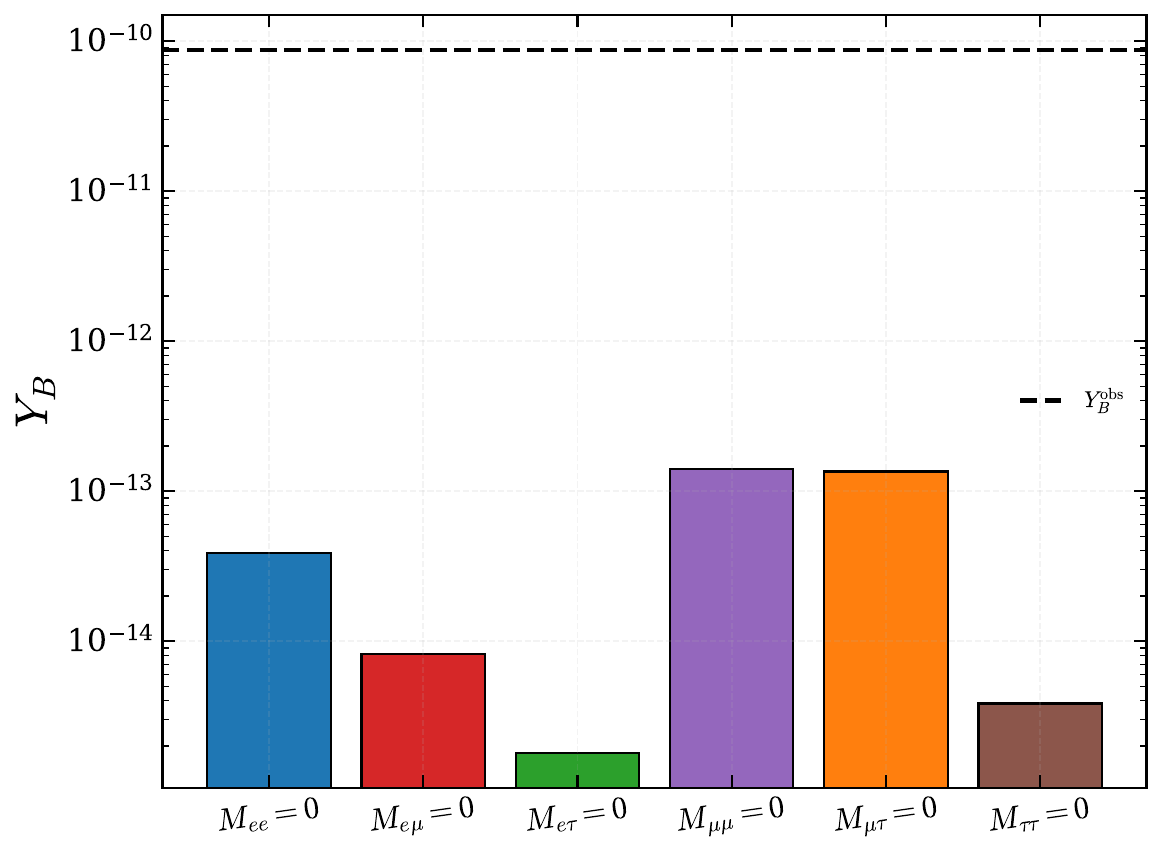}
\caption{
Baryon asymmetry generated in the hierarchical Type-II leptogenesis regime.}
\label{fig:hierarchical_YB}
\end{subfigure}
\hfill
\begin{subfigure}[t]{0.54\textwidth}
\centering
\includegraphics[width=\textwidth]
{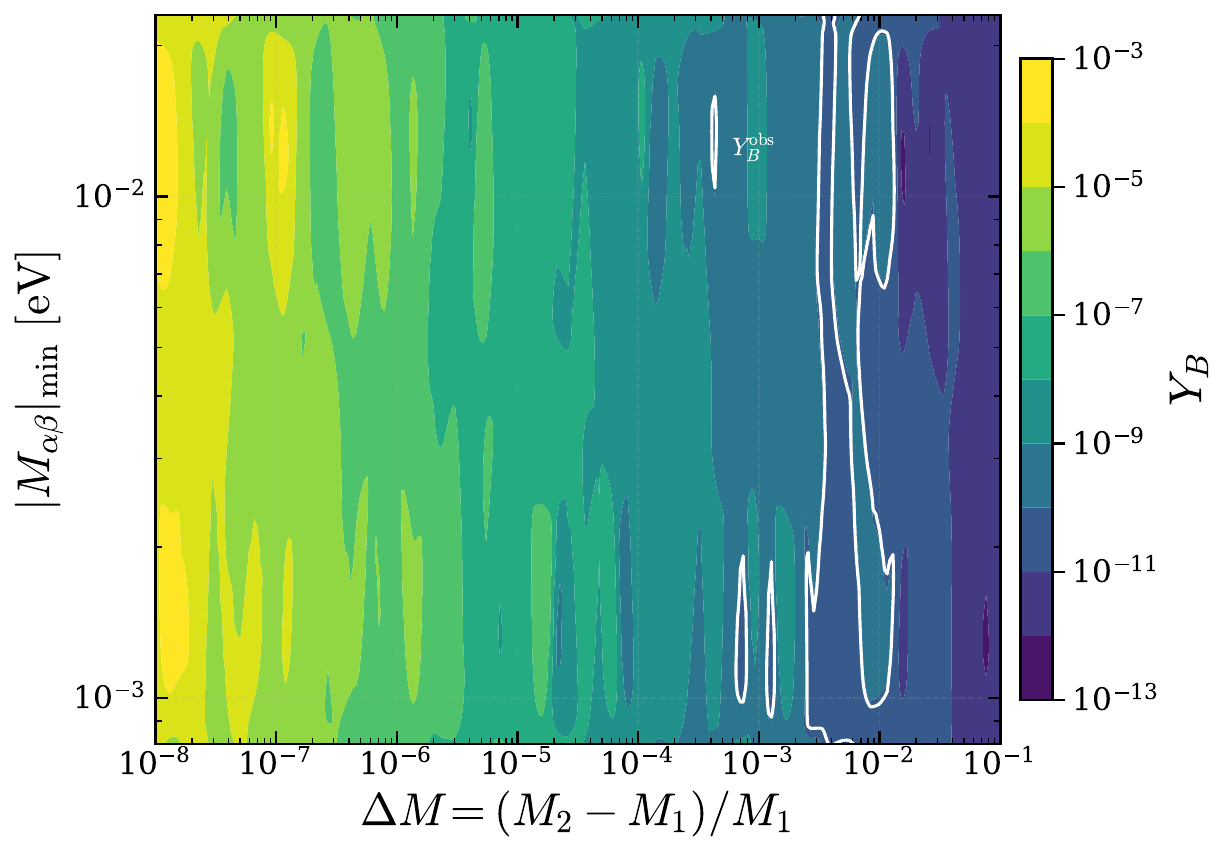}
\caption{
Combined phase-space analysis in the plane of the resonance parameter
$\Delta M=(M_2-M_1)/M_1$
and 
$|M_{\alpha\beta}|_{\rm min}$.
The white contour corresponds to $Y_B^{\rm obs}\simeq 8.7\times10^{-11}$.}
\label{fig:heatmap}
\end{subfigure}

\caption{
Comparison between hierarchical and resonant Type-II leptogenesis for the one-zero neutrino textures.
\textbf{(a)} In the hierarchical regime
($M_2\gg M_1$),
the generated baryon asymmetry remains strongly suppressed for all texture structures and fails to reproduce the observed cosmological value.
\textbf{(b)} In the resonant regime, the CP asymmetry becomes significantly enhanced as the mass splitting parameter
$\Delta M$
approaches zero.
A viable resonant manifold emerges in which successful baryogenesis and strong one-zero texture suppression
($|M_{\alpha\beta}|_{\rm min}\lesssim10^{-3}\ {\rm eV}$)
become simultaneously compatible.
The figure illustrates the central result of the present work: resonant enhancement can revive otherwise suppressed texture structures and generate the observed baryon asymmetry of the Universe.
}
\label{fig:hierarchical_vs_resonant}
\end{figure*}
%===========================================================
This suppression arises naturally from the neutrino mass constraints. Since the neutrino masses require comparatively small triplet Yukawa couplings, the corresponding CP asymmetry generated in the hierarchical regime becomes insufficient for successful baryogenesis.

The failure of hierarchical leptogenesis can be understood directly from the structure of Eq.~(\ref{eq:cp_asymmetry}). Without resonant enhancement, the generated asymmetry is limited by the small Yukawa couplings required to satisfy the neutrino mass constraints.

The corresponding baryon asymmetry obtained in the hierarchical regime is shown in Fig.~\ref{fig:hierarchical_YB}.

%===========================================================
\subsection{Phenomenological Implications}
%===========================================================

The results obtained in this work have several important phenomenological implications.

First, the analysis demonstrates that one-zero neutrino textures remain compatible with successful baryogenesis once resonant enhancement is included. This considerably enlarges the phenomenologically viable parameter space of Type-II seesaw models.

Second, the strong sensitivity of leptogenesis to flavor-dependent washout effects suggests that neutrino texture structures can leave observable imprints on thermal baryogenesis dynamics.

Third, the emergence of approximate universality in the strongly resonant regime indicates that the resonance mechanism itself becomes the dominant source of baryogenesis, partially washing out the differences among the texture structures.

Finally, the analysis highlights the importance of flavor-resolved thermal evolution in realistic leptogenesis studies. Since the one-zero textures intrinsically constrain the flavor structure of the Yukawa couplings, flavored washout effects become unavoidable in a consistent treatment of baryogenesis.
%

%===========================================================
\section{Conclusions}
%===========================================================

In this work we investigated the compatibility between one-zero neutrino mass textures and resonant Type-II leptogenesis within a two-triplet scalar framework. Our primary objective was to determine whether the flavor structures imposed by one-zero textures can simultaneously satisfy neutrino oscillation constraints and generate the observed baryon asymmetry of the Universe.

We first analyzed the structure of one-zero neutrino mass textures by reconstructing the Majorana neutrino mass matrix from the PMNS parameters. Extensive numerical scans over the oscillation parameter space were performed in order to classify the phenomenologically viable textures.

The numerical analysis revealed a hierarchical pattern among the one-zero structures. In particular, the off-diagonal textures, were found to be significantly easier to realize compared to the diagonal textures.

We then investigated the generation of CP asymmetry within the Type-II seesaw framework. Our analysis showed that hierarchical Type-II leptogenesis fails to produce the observed baryon asymmetry for the texture structures considered in this work. The resulting asymmetry remains strongly suppressed due to the small Yukawa couplings required by neutrino mass constraints.

The situation changes dramatically in the presence of two quasi-degenerate scalar triplets. In the resonant regime $M_2\simeq M_1$, the self-energy contribution becomes resonantly enhanced and the generated CP asymmetry increases substantially. We found that the baryon asymmetry exhibits an approximate inverse dependence on the mass splitting parameter,
\begin{equation}
Y_B
\propto
\frac{1}{\Delta M}.
\end{equation}

Our numerical scans demonstrated that successful baryogenesis becomes possible once the resonance condition is sufficiently satisfied. The resonant enhancement therefore provides an efficient mechanism for reviving otherwise suppressed one-zero texture structures.

To investigate the thermal origin of the asymmetry, we solved the Boltzmann equations governing the evolution of the scalar triplet abundance and the $B-L$ asymmetry. The thermal evolution revealed the expected freeze-out behavior: the triplet abundance initially tracks the equilibrium distribution before decoupling from the thermal bath, while the generated asymmetry approaches a constant plateau after washout processes become inefficient.

We further extended the analysis to a flavor-resolved framework in which the electron, muon, and tau asymmetries evolve independently. The flavored Boltzmann evolution demonstrated the existence of flavor-dependent washout effects leading to hierarchical asymmetry survival:
\begin{equation}
|Y_{\Delta_\mu}|
>
|Y_{\Delta_\tau}|
>
|Y_{\Delta_e}|.
\end{equation}

This establishes a direct connection between:
\begin{equation*}
\text{Neutrino Texture Geometry}
\longrightarrow
\text{Flavor Washout}
\longrightarrow
\text{Thermal Baryogenesis}.
\end{equation*}

The present work therefore demonstrates that one-zero neutrino textures can remain compatible with successful thermal leptogenesis once resonant enhancement and flavor-dependent dynamics are properly incorporated.

Our results highlight the importance of flavor geometry in determining the efficiency of baryogenesis and suggest that neutrino texture structures can play a significant role in shaping the thermal history of the early Universe.

Several important extensions remain possible. In particular, a more complete analysis would involve deriving the flavored CP asymmetries directly from the underlying texture Yukawa matrices, including full flavor-covariant kinetic equations and thermal corrections. Such studies could further clarify the interplay between neutrino flavor symmetries and thermal leptogenesis dynamics.

Nevertheless, the present analysis already establishes resonant Type-II leptogenesis as a viable framework in which one-zero neutrino textures, flavor-dependent washout effects, and thermal baryogenesis can be consistently realized.

\end{document}